\documentclass[]{spie}  

\usepackage{amsmath,amsfonts,amssymb}
\usepackage{graphicx}
\usepackage[colorlinks=true, allcolors=blue]{hyperref}
\usepackage{authblk}

\begin{document} 

\title{GRBAlpha and VZLUSAT-2: GRB observations with CubeSats after 3 years of operations}

\author[a]{Filip M\"{u}nz}
\author[a]{Jakub \v{R}ípa}
\author[b]{Andr\'as P\'al}
\author[a]{Marianna Daf\v{c}\'ikov\'a}
\author[a]{Norbert Werner}
\author[c]{Masanori Ohno}
\author[b]{L\'aszl\'o Mesz\'aros}
\author[d]{Vladimír Dániel}
\author[e]{Peter Han\'ak}
\author[f]{J\'an Hudec}
\author[f]{Marcel Frajt}
\author[f]{Jakub Kapu\v{s}}
\author[d]{Petr Svoboda}
\author[d]{Juraj Dud\'a\v{s}}
\author[k]{Miroslav Kasal}
\author[a]{Tom\'a\v{s} V\'itek}
\author[a]{Martin Kol\'a\v{r}}
\author[a]{Lea Szakszonov\'a}
\author[f]{Pavol Lipovsk\'y}
\author[a]{Michaela \v{D}ur\'i\v{s}kov\'a}
\author[g]{Ivo Ve\v{r}t\'at}
\author[d]{Martin Sabol}
\author[d]{Milan Junas}
\author[d]{Roman Maro\v{s}}
\author[a]{Pavel Kosík}
\author[l]{Zsolt Frei}
\author[c]{Hiromitsu Takahashi}
\author[c]{Yasushi Fukazawa}
\author[h]{G\'abor Galg\'oczi}
\author[b]{Balázs Csák}
\author[i]{Robert L\'aszl\'o}
\author[c]{Tsunefumi Mizuno}
\author[a]{Nikola Hus\'arikov\'a}
\author[j]{Kazuhiro Nakazawa}

\affil[a]{Masaryk University, Brno, Czech Republic}
\affil[b]{Konkoly Observatory, Budapest, Hungary}
\affil[c]{Hiroshima University, Japan}
\affil[d]{Czech Aerospace Research Center, Prague, Czech Republic}
\affil[e]{Technical University of Košice, Slovakia}
\affil[f]{Spacemanic Ltd, Brno, Czech Republic}
\affil[g]{University of West Bohemia, Pilsen, Czech Republic}
\affil[h]{Wigner Research Center for Physics, Budapest, Hungary}
\affil[i]{Needronix s.r.o., Bratislava, Slovakia}
\affil[j]{Nagoya University, Japan}
\affil[k]{Brno University of Technology, Czech Republic}
\affil[l]{E\"{o}tv\"{o}s Lor\'and University, Budapest, Hungary}
\authorinfo{E-mail: munz@physics.muni.cz}

\pagestyle{empty} 
\setcounter{page}{301} 

\makeatletter
\renewcommand*{\thetable}{\arabic{table}}
\renewcommand*{\thefigure}{\arabic{figure}}
\makeatother

\maketitle

\begin{abstract}
    GRBAlpha is a 1U CubeSat launched in March 2021 to a sun-synchronous LEO at an altitude of 550 km to perform an in-orbit demonstration of a novel gamma-ray burst detector developed for CubeSats. 
    VZLUSAT-2 followed ten months later in a similar orbit carrying as a secondary payload a pair of identical detectors as used on the first mission. 
    These instruments detecting gamma-rays in the range of 30-900 keV consist of a 56 cm2 5 mm thin CsI(Tl) scintillator read-out by a row of multi-pixel photon counters (MPPC or SiPM). 
    The scientific motivation is to detect gamma-ray bursts and other HE transient events and serve as a pathfinder for a larger constellation of nanosatellites that could 
    localize these events via triangulation. 

    At the beginning of July 2024, GRBAlpha detected 140 such transients, while VZLUSAT-2 had 83 positive detections, confirmed by larger GRB missions. 
    Almost a hundred of them are identified as gamma-ray bursts, including extremely bright GRB 221009A and GRB 230307A, detected by both satellites. 
    We were able to characterize the degradation of SiPMs in polar orbit and optimize the duty cycle of the detector system also by using SatNOGS radio network for downlink.
\end{abstract}


\section{INTRODUCTION}
\label{sec:intro}  

This year our GRBAlpha\cite{pal2023} satellite has completed 3 years in orbit, fulfilling its role of a pathfinder for a planned constellation of nanosatellites for gamma-ray burst (GRB) detection  and 
localization called CAMELOT\cite{werner2018}. It was followed by other CubeSats, like SpIRIT\cite{spirit2023} (featuring a HERMES-SP gamma detector), EIRSAT-1\cite{dunwoody2023} or BurstCube\cite{racusin2017}, demonstrating an ongoing revolution of small satellites 
complementing large missions in high-energy astrophysics. Dominant approach that should allow all-sky coverage of gamma-ray transients is to measure shift in arrival time using 
cross-correlation of lightcurves, hopefully reaching
tenth of a millisecond precision needed for detectors distributed on low Earth orbits.

Although the scintillator-based detector of GRBAlpha is designed to be rather simple and robust compared to delicate instruments like HERMES
, it has now accumulated very nice statistics of GRB detections and proved long-term performance evolution of SiPM detectors. The same device (in two copies) was placed as a 
secondary payload on a 
3U CubeSat designed by Czech Aerospace Research Center (VZLU) called VZLUSAT-2\cite{Daniel2020} that is (in contrast with GRBAlpha) equipped with an active attitude control. It has however limitations in power and data-transfer budgets so 
that our gamma-ray detectors can operate only cca 1/3 of time.

Finally, a new 2U satellite named GRBBeta was integrated this year and is ready to be launched with Arianne 6 first flight. On the detector side, there are only minor improvements of the read-out 
electronics, however the satellite is equipped with S-band datalink that (together with AOCS system) should allow download data sampled at much higher rate.

Last but not least, we profit of the fact that the payload software (including FPGA code) can be updated on board. A major change in encoding and data transmission scheme, that was performed on GRBAlpha in late 2022, 
allowed to increase duty cycle to almost 100\% (in case of a flawless telecommanding done twice a day from our principal station in Ko\v{s}ice, Slovakia). Using amateur UHF frequencies for both uplink
and downlink, we could include some of our other stations into the SatNOGS network\footnote{\url{http://satnogs.org}} of radio stations and to use other selected stations of the network for data dropping
of pre-selected chunks of measurements that correspond to times of events reported elsewhere. 
Thanks to this, statistics of detected gamma-ray events increased ten-fold since the upgrade. 

\section{COMMUNICATION}

Since the failure of the VHF radio module onboard (November 2021) uplink has to rely on a noisier UHF band. Current solution uses one simplex station (Technical University, Košice) for transmission 
and reception of telemetry is forwarded from Piszk\'estet\H{o} Observatory, Hungary, or Jablonec, Slovakia, also in a simplex mode. Recently, with a new radio license a station
located at Konkoly Observatory in Budapest, Hungary, became available as a backup for telecommanding. Planning acquisition and downlink is simple thanks to a CubeSat control terminal called {\ttfamily vcom} 
developed by VZLU (with contribution of Spacemanic and our Hungarian colleagues). Distinct components on board have their proper set of sub-commands that are gradually evolving to offer more 
advanced features. For example, an ID of a ground station in the SatNOGS database and a (fractional) position in the data stored on board (one ``file'' corresponds to roughly 12hrs of measurement) 
is sufficient to plan a data drop: the software will retrieve station position, get times of next contact, plan follow-up of GRBAlpha with this station and schedule a data transmission on the payload's 
control unit. Thanks to a versatile design of the on-board communication (using either I2C or CAN protocols) payload electronics can directly send data to the radio transmitter.

Currently, data are saved in a self-synchronizing variable-length code that besides providing efficient compression of Poissonian data (in many channels dominated by small numbers 
or whole zero blocks) it also allows bit-level identification of stored data structures (mostly spectra, temperatures and timestamps). For more details, see A.P\'al 2023\cite{pal2023}.
During transmission, the selected block of data (a few thousand seconds around the time of the event), too big to fit in as single radio packet, is split into chunks (typically of 128 bytes)
and dropped in random order to selected ground stations. Some 15\% redundancy in data dropping should allow to reconstruct the original lightcurve taking into account 
usual packet loss during transmission fading.

\begin{figure} \label{fg:satnogs}
    \begin{center}
    \includegraphics[width=0.7\linewidth]{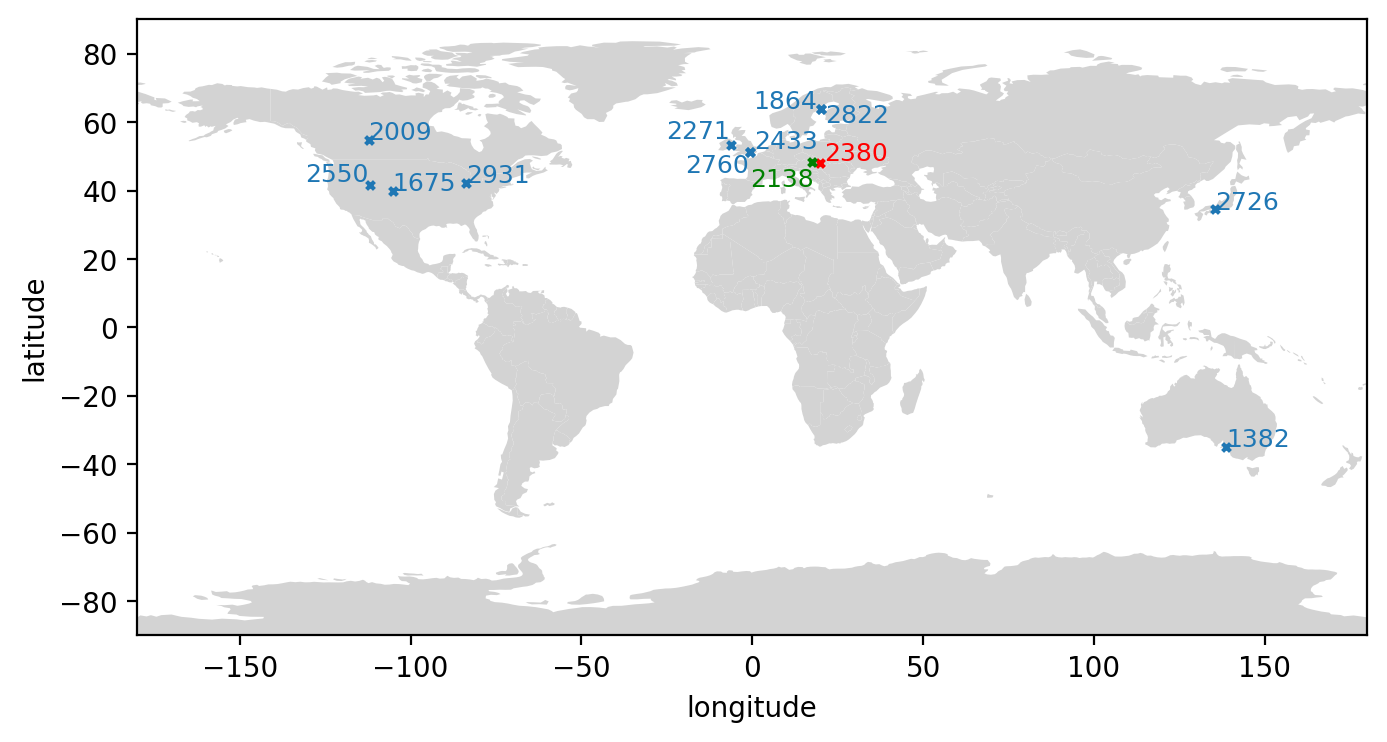}
    \end{center}
    \caption[short]{Selected radio stations most used for data dropping of GRBAlpha. Piszk\'estet\H{o} station is marked red, secondary on in Jablonec in green. Credit M. Dafčíková, MSc thesis.}
\end{figure}

\section{DETECTOR PERFORMANCE}
\label{sec:perform}  

Each detector consists of a CsI(Tl) crystal scintillator wrapped in an ESR foil with one narrow side open where a pair of 4 SiPMs (S13360-3050PE by Hamamatsu, 3$\times$3 mm active area each) are glued.
Aluminum box (1mm thick) was not light-tight enough so a DuPont TCC15BL3 polyvinyl fluoride (PVF) tedlar had to be used as extra wrapping of the setup. Photodetectors were protected from protons with  
2.5~mm thick PbSb alloy shield (adding a substantial mass to the weight of the whole device).

Combined signal of 4 SiPMs of each channel is then amplified and shaped in an analog part of the electronic board, reaching characteristic pulse width of 15~$\mu$s. Then
the signal is digitized (at cca 600 MHz sampling) and pulse amplitudes are stored in a histogram by FPGA. Each channel having its pair of FPGA and MCU, these 256-bin spectra
are read out at chosen cadence and stored with required binning in a memory within the same digital board.\footnote{Two channel design provides necessary redundancy, however,
only one channel operates at given time.}

\subsection{Radiation Degradation}
\label{ssec:gain}  

For housekeeping purposes we collect 60s spectra with full (256-bin) resolution usually several times a week. 
Noise peak (originally from ground calibrations around channel 40 in full resolution spectra) is showing gradual shift to higher energies -- see fig. 2 -- result of damage to MPPC by cosmic radiation 
(evident correlation with solar activity and CMEs). Increasing particle densities in Earth's magnetosphere (as we approach maximum of the Sun cycle) are slightly balanced by orbital degradation 
(altitude decreased by 50~km in 3 years): detailed account combining data 
from both satellites (incl. comparison with cosmic ray background models) will be published in a near future. 

\begin{figure} \label{fg:degrad}
	\begin{center}
    \includegraphics[width=0.7\linewidth]{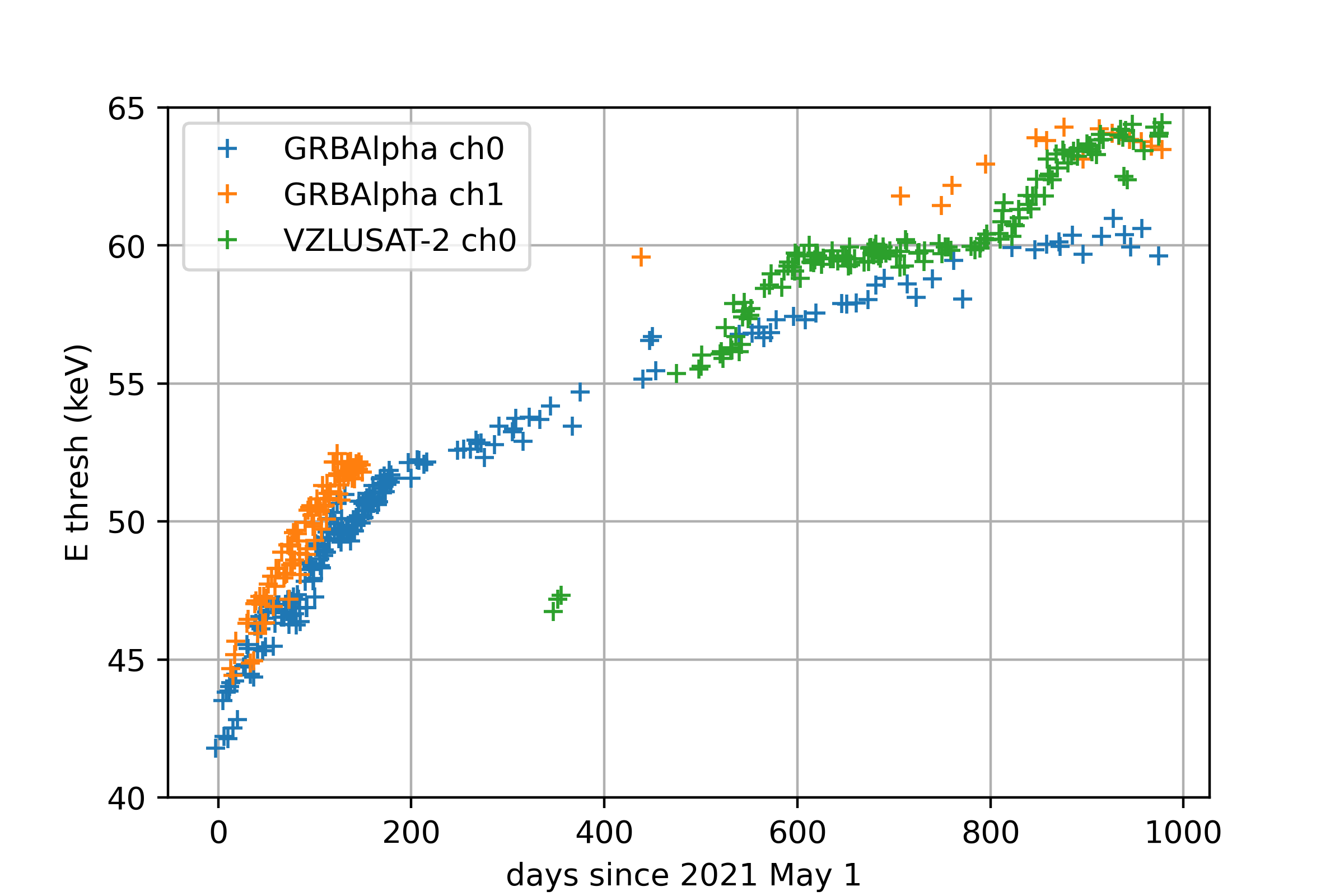}
    \end{center}
    \caption{Detector threshold evolution (based on noise peak degradation) through most of the lifetime of GRBAlpha. Errorbars correspond to the spread of values from multiple spectra
    measured around a given date; it can be partially attributed to temperature variations throughout the orbit.}
\end{figure}        

\begin{figure}
    \begin{center}
        \includegraphics[height=32ex]{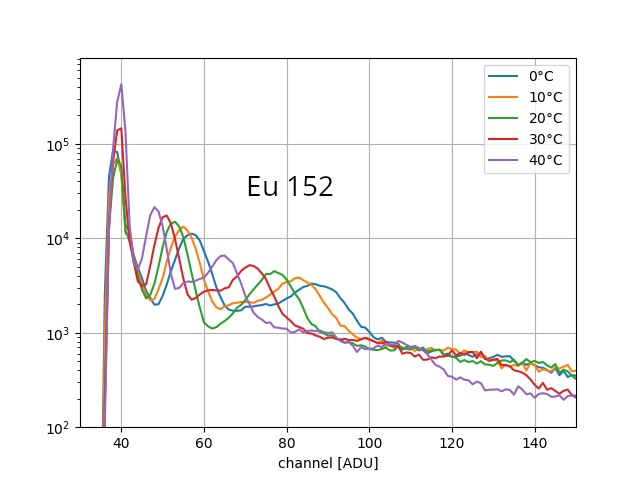} \par 
        \  \includegraphics[height=36ex]{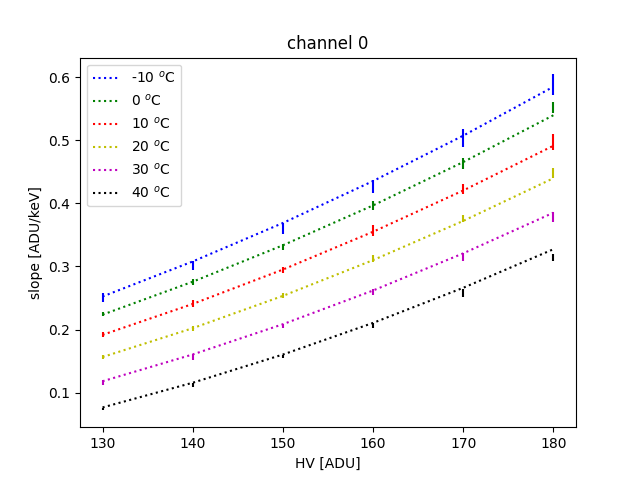}
        \  \includegraphics[height=36ex]{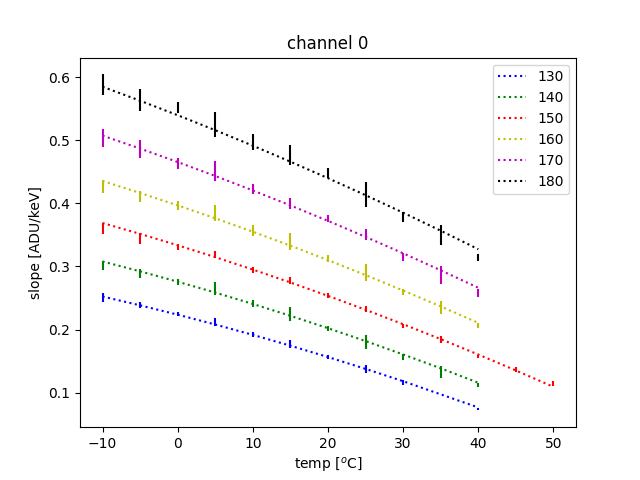}
\end{center}
    \label{fg:tempcal}
    \caption[short]{Peak position (measured in ADC channels) and gain variations with respect to the detector temperature and operating voltage (given in DAU units). Temperature is 
    measured at 3 points of the MPPC board (center and both edges, average is taken as a reference); thanks to good heat conductivity of the aluminum casing we assume the temperature of bulk of the scintillator 
    should be close to this value. Dotted curves follow formula in eq. \ref{eq:calquad}.}
    \end{figure}

\subsection{Gain Calibration}
\label{ssec:gain}  

Advantage of MPPCs compared to traditional photomultipliers is much lower operational voltage, some 3-5 volts above breakdown voltage, usually 48-51 volts. 
Following calibration tests were performed with detectors assembled for GRBBeta satellite. Four sets of 4 SiPMs were pre-selected (out of cca 40 MPPCs provided from Hamamatsu) 
to make as uniform group as possible with respect to the (factory measured) breakdown voltages and dark currents. The groups with the smallest spread were used for
a board selected as a flight model, next to it were MPPC groups used for a flight-spare detector. This one was (after successful integration of the satellite) equipped with
a universal power supply (one board of 10$\times$10~cm CubeSat footprint) and a spare electronic board and undertook a series of (temperature-controlled) calibration measurements with different gamma-ray sources both at Masaryk University, Brno, and
at laboratories of ADVACAM company, Prague. The spectral resolution of current detector design limits usability of certain isotopes with too high concentration of spectral
lines. Complete setup with a water cooling block was fitted in a 30$\times$20~cm sealed plastic container (allowing N2 purging to prevent any water condensation when 
measuring below the freezing point). Spectra obtained for Eu152 isotope and fitted calibration curves are shown on fig. 3. Distance from the noise peak (mentioned above)
measured in ADU (channel number) scales to a good precision linearly with deposited energy, however this scaling depends both on operation voltage $V$ and temperature $T$ 
of the device. The voltage, moreover, is set through a custom digital-analog converter so for practical purposes $V$ is measured in DAU units, not in volts.\footnote{Conversion
can be calculated with a formula $V[\mathrm{V}] = 48.8 + 0.0612\ V[\mathrm{DAU}]$ (values given for channel 0 of GRBAlpha).}

\newcommand{\K}{\mathrm{K}}
\newcommand{\maa}{{\mathrm{a}}}

	Channel position was finally modelled as (temp. $T$ in ${}^\circ$C, $V$ in instr. DAU units) 
    \begin{multline}
        \chi = 39.9 + E/MeV \times \\
        \quad \left ( 334 + 6.05 (V-150) - T\ 3.7 \K^{-1} + 0.027 (V - 150)^2 - T^2\ 0.016 \K^{-2} - (V-150)\ T\ 0.033 \K^{-1}\right ) \label{eq:calquad}
\end{multline}
    According to the manufacturer's datasheet MPPC gain should decrease with temperature in a manner that overrides small increase of the light yield of the CsI scintillator.
    This calibration should serve to design adjustment procedure of operating voltage to compensate temperature variations throughout the orbit.  In the case of GRBAlpha,
    where the detector is exposed directly to the sunlight, the temperature varies (with orbital period) from 0 to 18 ${}^\circ$C, detectors on VZLUSAT-2 are
    on average 7${}^\circ$C colder. Continuous rotation of GRBAlpha should to some degree smear out these variations while for GRBBeta stable attitude can in some cases
    result in much higher detector temperatures: here the gain variation should be compensated regularly not to distort collected scientific data.

\section{DETECTIONS}

\newcommand{\stats}{
			  \begin{tabular}{l r r c}
				\textbf{satellite} & \textbf{GRBAlpha} & \textbf{VZLUSAT-2}  \\
                \hline
				long GRBs & 67 & 41  \\
				short GRBs & 14 & 3  \\
				solar flares & 57 & 34  \\
				SGRs & 2 & 3 \\
				total & 140 & 83 \\
			  \end{tabular}
}
\begin{table}
    \centering \stats
    \label{tb:detect}.
    \caption{Statistics of significant detections as for May 2024. }
    \end{table}

As already mentioned above, since upgrade of the on-board software (21 months ago) the duty cycle for GRBAlpha satellite increased significantly and the detection rate (in coincidence with external trigger sources) 
has risen up to roughly 1.6 gamma-ray event per week. The comprehensive table for identified events (with links to related notices, graphs and data) are available at 
\url{https://monoceros.physics.muni.cz/hea/GRBAlpha/} or \url{https://monoceros.physics.muni.cz/hea/VZLUSAT-2/} for GRBAlpha
or VZLUSAT-2 detections, respectively. Occurrences of different types of detections are given in table \ref{tb:detect} (identifications come from reports of observations of
other missions and in other spectral domains, mostly distributed through GCNs\footnote{General Coordinates Network, \url{https://gcn.nasa.gov/}}).
    
The most remarkable detection, that merited its own article\cite{ripa2023}, was GRB 221009A, nicknamed BOAT meaning ``brightest of all times'': flux from this burst even disturbed
Earth's ionosphere. Major GRB observatories like FERMI/GBM or Konus/Wind were saturated while GRBAlpha experienced only a moderate pile-up which allowed us to measure directly the
gamma peak rate. Without attitude control (and knowledge) we cannot convert this rate to corresponding flux. However, we were fortunate to collect data in 13 spectral bands at that
moment and spectral reconstruction allowed us to approximate most probable direction of the source. The reported peak flux (averaged over 4s time bins) is 8.4$\times 10^{52}$ erg/s, 
assuming a curved power-law model for the spectrum (its shape was determined outside the region affected by pile-up).

Second such event, GRB230307A, was luckily detected by both satellites\footnote{At these altitudes cca 30\% of the sky is occulted by Earth so probability of bare visibility by 2 satellites is quite high.}
the lightcurves were reported in GCNs 33418\cite{2023GCN.Alpha} and 33424\cite{2023GCN.VZL}. Sampling rates were still far too low to attempt any cross-correlation for triangulation
purposes but this very bright event, candidate for a kilonova progenitor, is still worth of a detailed study.

\section{BACKGROUND MONITORING}
\label{sec:bkg}

\begin{figure} \label{fg:bkgmaps}
    \begin{tabular}{l r}
\includegraphics[width=0.45\linewidth]{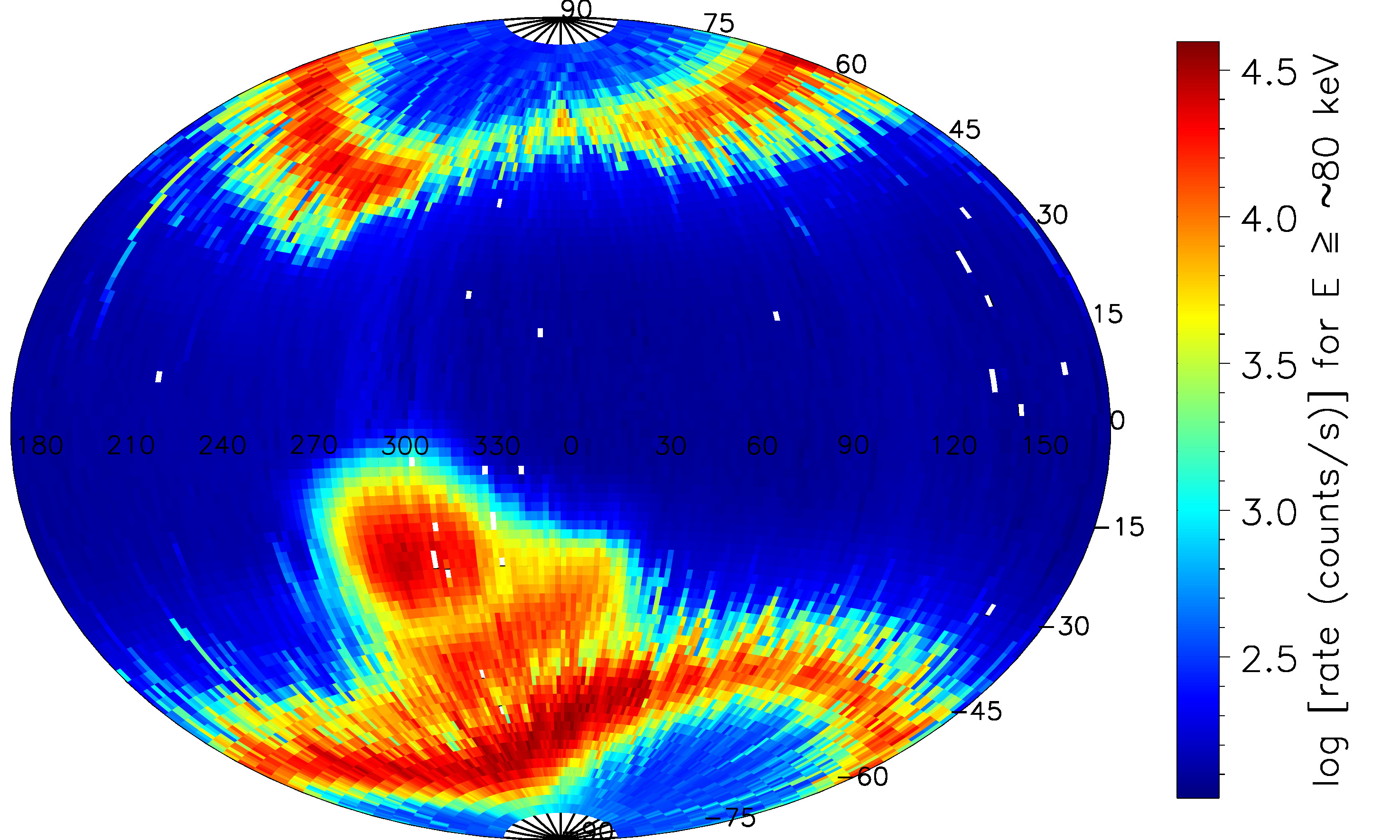} &
\includegraphics[width=0.45\linewidth]{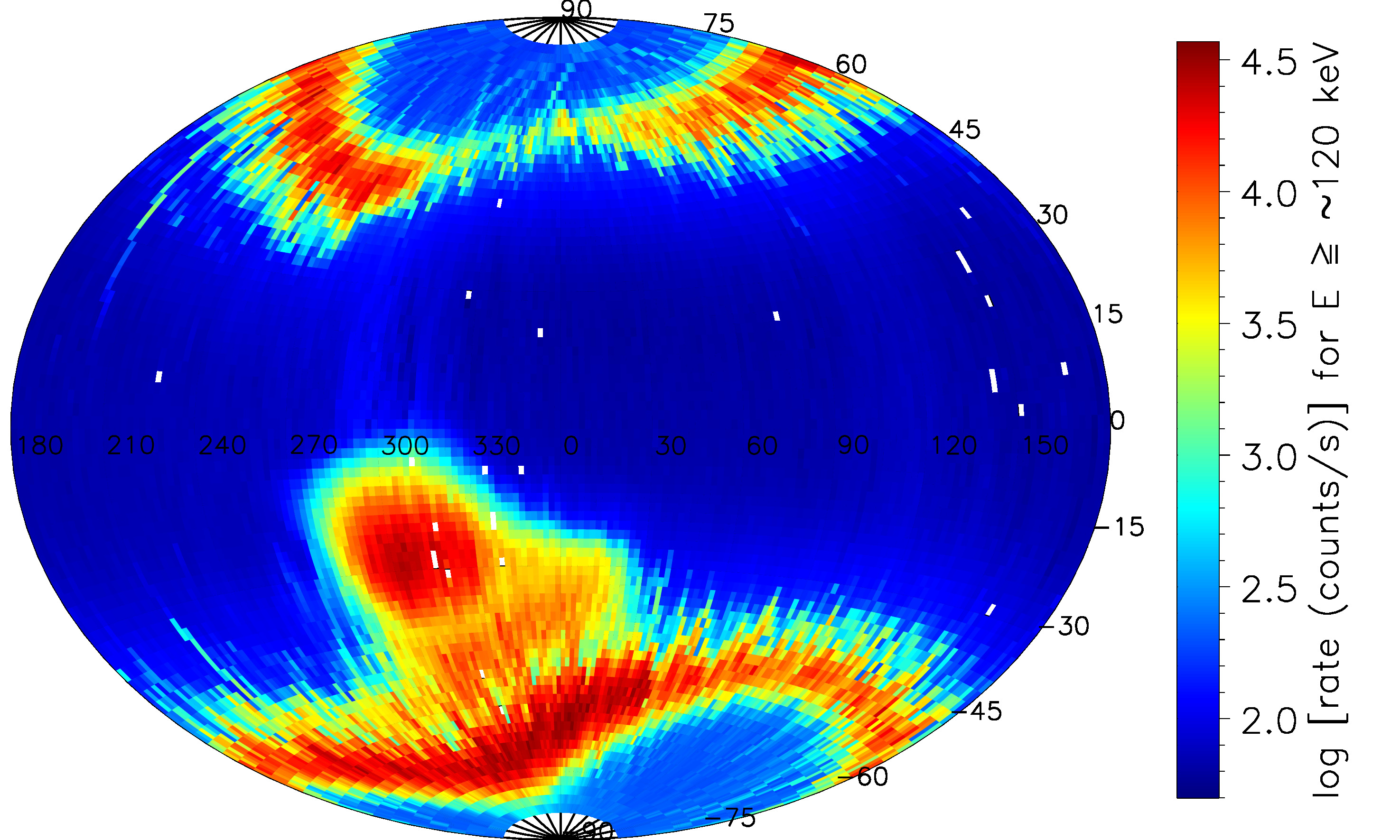} \\
\includegraphics[width=0.45\linewidth]{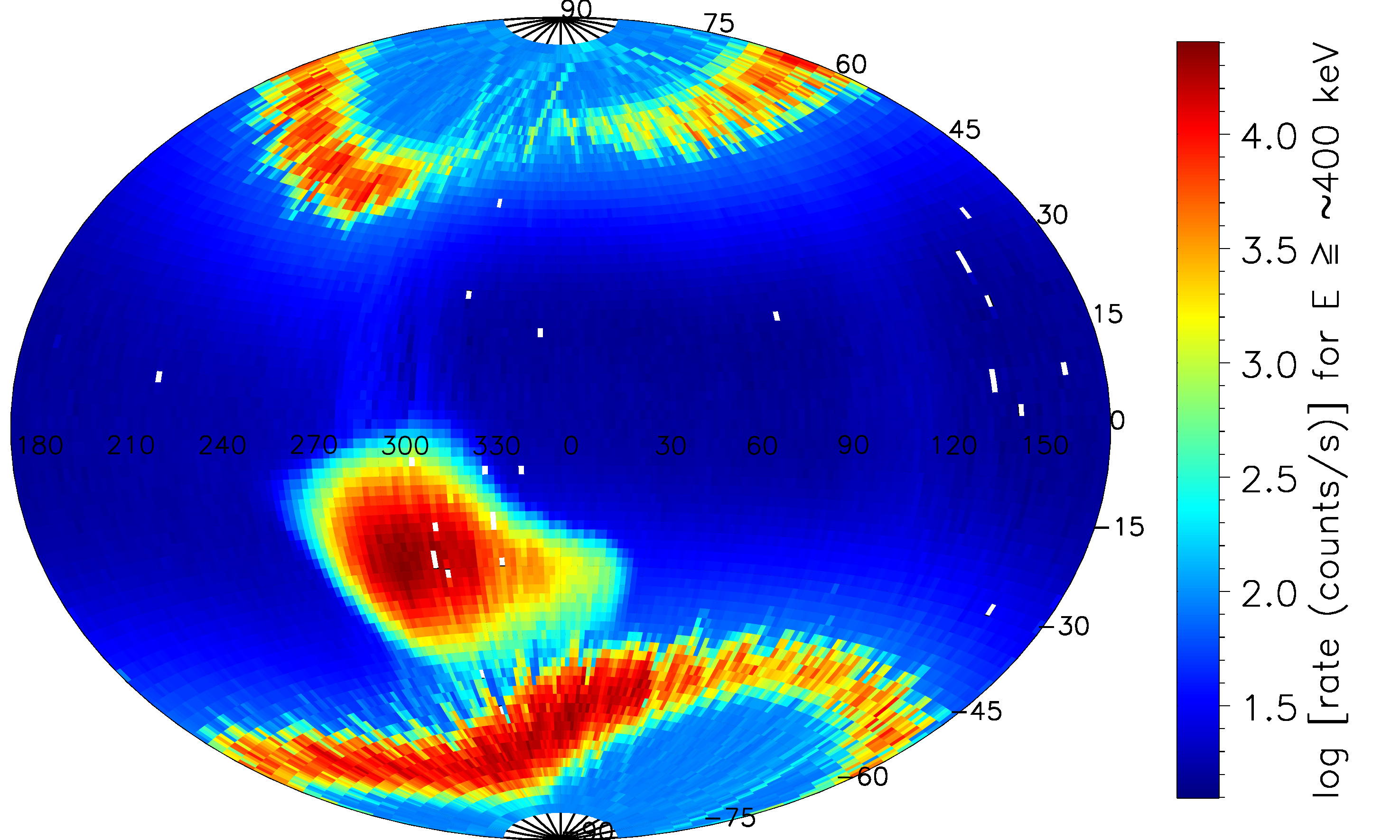} &
\includegraphics[width=0.45\linewidth]{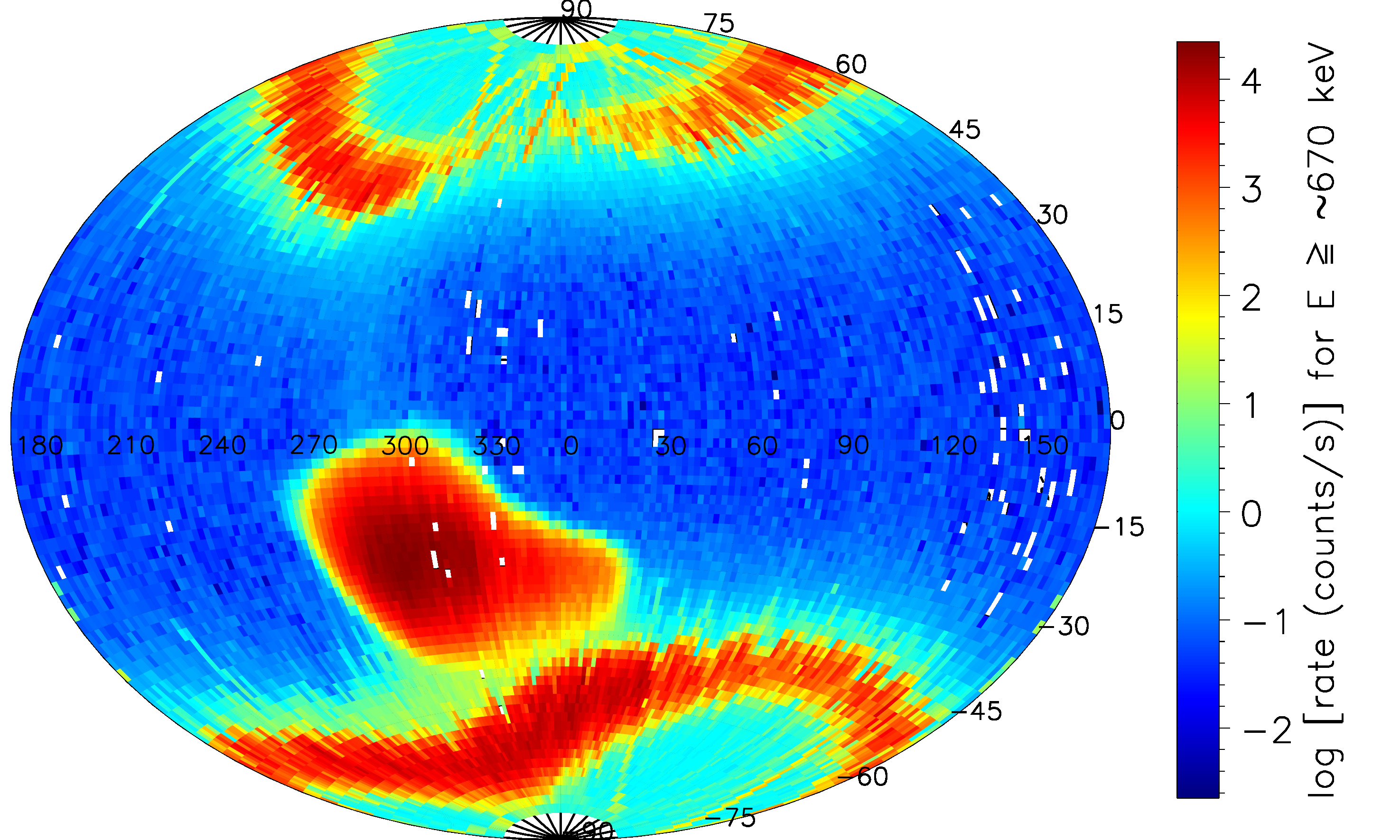} \\
\end{tabular}
\caption[short]{Background maps from data collected from Nov 2022 to Jul 2023, with increasing low-energy threshold.}
\end{figure}

Amount of data downloaded with new downlink possibilities allows constructing background maps with much denser coverage, reducing need for interpolation and revealing finer structures in 
particle background maps. Secular events (like coronal mass ejections) have clear influence on the size of polar belts  -- feature that could be easily recorded using our detector. We download roughly 
110-120 hrs of 4-channel lightcurves per month (with 2~Hz sampling rate), which corresponds to cca 75 complete orbits. In fig. \ref{fg:bkgmaps} you find maps from a considerably
larger dataset collected through approximately 10 months of observations. Ragged contours of polar belts correspond probably to the above-mentioned variations of spread of belts and SAA during this 
period.

\begin{figure} \label{fg:longline}
    \begin{center}
        \includegraphics[height=30ex]{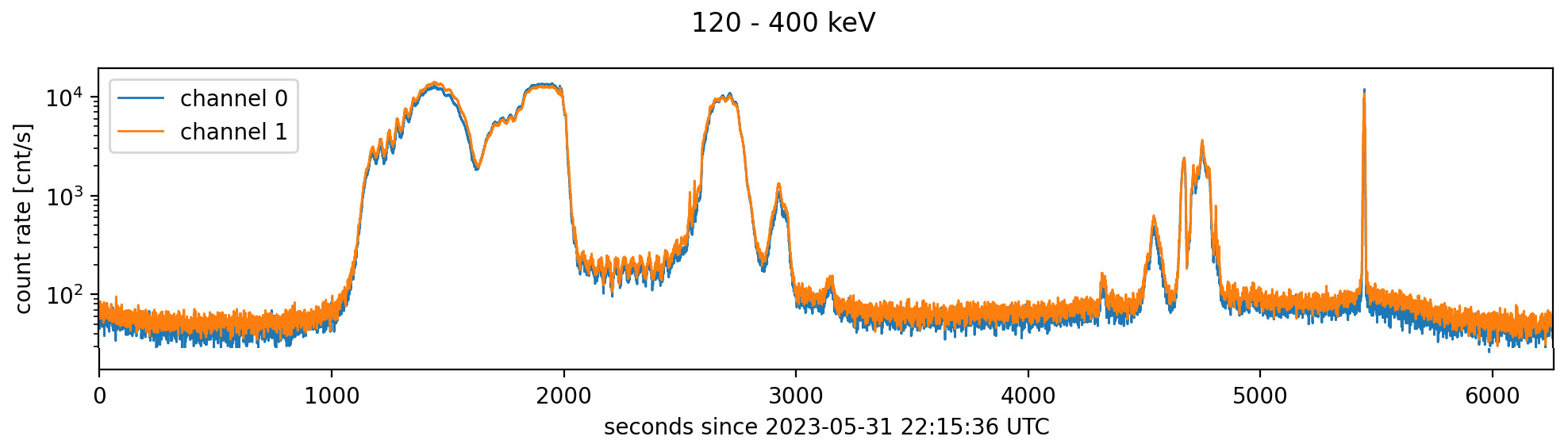} 
    \end{center}
    \caption[short]{Data measured by GRBAlpha in 2 channels along entire orbit, passing through South Atlantic Anomaly, southern and northern polar rings.
    Periodic signal due to motion of the satellite are visible around time 1300~s and 2200~s.}
\end{figure}

Due to a wobbling motion
of GRBAlpha satellite, there are also observations of periodic signal as seen on figure \ref{fg:longline}. This could be explained assuming some level of anisotropy in 
particle background. Observations (mostly in polar regions) are so far too sparse to assert whether this is a permanent or a transient feature. Apart from this, regions 
around poles exhibit higher but predictable background level that with proper models should allow to implement a trigger algorithm with a good specificity (i.e. low false
alert rate) in this part of the orbit.


\section{Summary}

With a remarkable number of GRBs and other events registered with rather simple gamma detectors on board of GRBAlpha and VZLUSAT-2 we gather a strong support for the feasibility
of a nanosatellite constellation covering high-energy transient sky. We also gradually improve understanding of our device and its performance over time spent in orbit. 
Background monitoring on polar LEOs is another outcome of these missions that are likely to maintain their full capabilities until the end of their orbital lifetime.
In the case of VZLUSAT-2, other instruments on board\cite{Lobste2022} can provide simultaneous data in softer band to create more comprehensive picture of local space environment.
And with upcoming launch of a GRBBeta 2U CubeSat (and its much higher downlink capacities) we can finally get closer to the precise timing needed for direction reconstruction of gamma-ray bursts.

\acknowledgments 
 
This contribution was mainly funded by Czech Science Foundation (GA\v{C}R) grant 24-11487J. 
We acknowledge support by the grants KEP2/2020 and SA-40/2021 of the Hungarian Academy of Sciences
and E\"otv\"os Lor\'and Research Network, respectively, for satellite components and payload developments and the
grant IF-7/2020 for providing the financial support for ground infrastructure. We acknowledge supported by
the MUNI Award for Science and Humanities funded by the Grant Agency of Masaryk University. The research
leading to these results has received funding from the European Union’s Horizon 2020 Program under the
AHEAD2020 project (grant agreement n. 871158). We are grateful to the operators and developers of the
SatNOGS network for providing a
robust framework for bulk data download from our detectors flying above radio stations involved. 

\bibliographystyle{spiebib} 

\bibliography{poster.bib} 

\begin{thebibliography}{10}

\bibitem{pal2023}
Pál, A., Ohno, M., Mészáros, L., Werner, N., Řípa, J., Csák, B., Dafčíková, M., Frajt, M., Fukazawa, Y., Hanák, P., Hudec, J., Husáriková, N., Kapuš, J., Kasal, M., Kolář, M., Koleda, M., Laszlo, R., Lipovský, P., Mizuno, T., Münz, F., Nakazawa, K., Rezenov, M., Šmelko, M., Takahashi, H., Topinka, M., Urbanec, T., Riffald Souza~Breuer, J.~B., Bozóki, T., Dálya, G., Enoto, T., Frei, Z., Friss, G., Galgóczi, G., Hroch, F., Ichinohe, Y., Kapás, K., Kiss, L.~L., Matake, H., Odaka, H., Poon, H., Povalač, A., Takátsy, J., Torigoe, K., Uchida, N., and Uchida, Y., ``{GRBAlpha: The smallest astrophysical space observatory : I. Detector design, system description, and satellite operations},'' {\em Astronomy \& Astrophysics}~{\bf 677},  A40 (2023).

\bibitem{werner2018}
{Werner}, N., {{\v{R}}{\'\i}pa}, J., {P{\'a}l}, A., {Ohno}, M., {Tarcai}, N., {Torigoe}, K., {Tanaka}, K., {Uchida}, N., {M{\'e}sz{\'a}ros}, L., {Galg{\'o}czi}, G., {Fukazawa}, Y., {Mizuno}, T., {Takahashi}, H., {Nakazawa}, K., {V{\'a}rhegyi}, Z., {Enoto}, T., {Odaka}, H., {Ichinohe}, Y., {Frei}, Z., and {Kiss}, L., ``{{CAMELOT: Cubesats Applied for MEasuring and LOcalising Transients mission overview}},'' ~{\bf 10699},  106992P (July 2018).

\bibitem{spirit2023}
Thomas, M., Trenti, M., Sanna, A., Campana, R., Ghirlanda, G., Řípa, J., Burderi, L., Fiore, F., Evangelista, Y., Amati, L., Barraclough, S., Auchettl, K., del Castillo, M.~O., Chapman, A., Citossi, M., Colagrossi, A., Dilillo, G., Deiosso, N., Demenev, E., Longo, F., Marino, A., McRobbie, J., Mearns, R., Melandri, A., Riggio, A., Di~Salvo, T., Puccetti, S., and Topinka, M., ``Localisation of gamma-ray bursts from the combined spirit plus hermes-tp/sp nano-satellite constellation,'' {\em Publications of the Astronomical Society of Australia}~{\bf 40} (2023).

\bibitem{dunwoody2023}
Dunwoody, R., Doyle, M., Murphy, D., Finneran, G., O’Callaghan, D., Reilly, J., Thompson, J., Akarapu, S. K.~R., de~Barra, C., Cotter, L., et~al., ``{Development, description, and validation of the operations manual for EIRSAT-1, a 2U CubeSat with a gamma-ray burst detector},'' {\em Journal of Astronomical Telescopes, Instruments, and Systems}~{\bf 9}(3),  037001--037001 (2023).

\bibitem{racusin2017}
Racusin, J., Perkins, J.~S., Briggs, M.~S., de~Nolfo, G., Krizmanic, J., Caputo, R., McEnery, J.~E., Shawhan, P., Morris, D., Connaughton, V., Kocevski, D., Wilson-Hodge, C., Hui, M., Mitchell, L., and McBreen, S., ``{BurstCube: A CubeSat for Gravitational Wave Counterparts},'' (2017).

\bibitem{Daniel2020}
{D{\'a}niel}, V., {Svoboda}, P., {Junas}, M., {Sabol}, M., {Cag{\'a}{\r{A}}}, J., {Stejskal}, M., {Dudas}, J., {Mikulickova}, L., {Marek}, A., {Pavlica}, R., {Sedl{\'a}{\v{c}}kov{\'a}}, P., and {Jech}, D., ``{{Development of CubeSat with COTS camera enabling EO with high GSD}},'' in [{\em Sensors, Systems, and Next-Generation Satellites XXIV}{\nolinebreak\hspace{0.1em}]},  {Neeck}, S.~P., {H{\'e}li{\`e}re}, A., and {Kimura}, T., eds., {\em Society of Photo-Optical Instrumentation Engineers (SPIE) Conference Series} {\bf 11530},  115300Z (Sept. 2020).

\bibitem{ripa2023}
{{\v{R}}{\'\i}pa}, J., {Takahashi}, H., {Fukazawa}, Y., {Werner}, N., {Munz}, F., {Pal}, A., {Ohno}, M., {Dafcikova}, M., {Meszaros}, L., {Csak}, B., {Husarikova}, N., {Kolar}, M., {Galgoczi}, G., {Breuer}, J.-P., {Hroch}, F., {Hudec}, J., {Kapus}, J., {Frajt}, M., {Rezenov}, M., {Laszlo}, R., {Koleda}, M., {Smelko}, M., {Hanak}, P., {Lipovsky}, P., {Urbanec}, T., {Kasal}, M., {Povalac}, A., {Uchida}, Y., {Poon}, H., {Matake}, H., {Nakazawa}, K., {Uchida}, N., {Bozoki}, T., {Dalya}, G., {Enoto}, T., {Frei}, Z., {Friss}, G., {Ichinohe}, Y., {Kapas}, K., {Kiss}, L.~L., {Mizuno}, T., {Odaka}, H., {Takatsy}, J., {Topinka}, M., and {Torigoe}, K., ``{{The peak-flux of GRB 221009A measured with GRBAlpha}},'' {\em arXiv e-prints}~{\bf 677},  L2 (Feb. 2023).

\bibitem{2023GCN.Alpha}
{Dafcikova}, M., {Ripa}, J., {Pal}, A., {Werner}, N., {Ohno}, M., {Takahashi}, H., {Meszaros}, L., {Csak}, B., {Husarikova}, N., {Munz}, F., {Topinka}, M., {Kolar}, M., {Breuer}, J.~P., {Hroch}, F., {Urbanec}, T., {Kasal}, M., {Povalac}, A., {Hudec}, J., {Kapus}, J., {Frajt}, M., {Laszlo}, R., {Koleda}, M., {Smelko}, M., {Hanak}, P., {Lipovsky}, P., {Galgoczi}, G., {Uchida}, Y., {Poon}, H., {Matake}, H., {Uchida}, N., {Bozoki}, T., {Dalya}, G., {Enoto}, T., {Frei}, Z., {Friss}, G., {Fukazawa}, Y., {Hirose}, K., {Hisadomi}, S., {Ichinohe}, Y., {Kapas}, K., {Kiss}, L.~L., {Mizuno}, T., {Nakazawa}, K., {Odaka}, H., {Takatsy}, J., {Torigoe}, K., {Kogiso}, N., {Yoneyama}, M., {Moritaki}, M., {Kano}, T., and {GRBAlpha Collaboration.}, ``{{GRB 230307A: Detection by GRBAlpha}},'' {\em {GRB Coordinates Network}}~{\bf 33418},  1 (Mar. 2023).

\bibitem{2023GCN.VZL}
{Ripa}, J., {Dafcikova}, M., {Pal}, A., {Werner}, N., {Ohno}, M., {Meszaros}, L., {Csak}, B., {Takahashi}, H., {Munz}, F., {Topinka}, M., {Hroch}, F., {Husarikova}, N., {Breuer}, J.~P., {Hudec}, J., {Kapus}, J., {Frajt}, M., {Rezenov}, M., {Laszlo}, R., {Galgoczi}, G., {Uchida}, N., {Enoto}, T., {Frei}, Z., {Fukazawa}, Y., {Hirose}, K., {Matake}, H., {Hisadomi}, S., {Ichinohe}, Y., {Kiss}, L.~L., {Mizuno}, T., {Nakazawa}, K., {Odaka}, H., {Torigoe}, K., {Svoboda}, P., {Daniel}, V., {Dudas}, J., {Junas}, M., {Gromes}, J., {Vertat}, I., and {VZLUSAT-2/GRB Payload Collaboration.}, ``{{GRB 230307A: VZLUSAT-2 detection}},'' {\em {GRB Coordinates Network}}~{\bf 33424},  1 (Mar. 2023).

\bibitem{Lobste2022}
{Granja}, C., {Hudec}, R., {Mar{\v{s}}{\'\i}kov{\'a}}, V., {Inneman}, A., {P{\'\i}na}, L., {Doubravova}, D., {Matej}, Z., {Daniel}, V., and {Oberta}, P., ``{{Directional-Sensitive X-ray/Gamma-ray Imager on Board the VZLUSAT-2 CubeSat for Wide Field-of-View Observation of GRBs in Low Earth Orbit}},'' {\em Universe}~{\bf 8},  241 (Apr. 2022).

\end{thebibliography}

\end{document}